# Ultrafast tristable spin memory of a coherent polariton gas

R. Cerna[1], Y. Léger[2*], T. K. Paraïso[3], M. Wouters[4], F. Morier-Genoud[1], M. T. Portella-Oberli[1] and B. Deveaud[1]

[1] Laboratory of Quantum Optoelectronics LOEQ, EPFL, CH-1015 Lausanne, Switzerland
[2] FOTON Laboratory, Université Européenne de Bretagne, CNRS-INSA-UR1, INSA de Rennes F-35708 Rennes, France
[3] Thomas J. Watson, Sr., Laboratory of Applied Physics, California Institute of Technology, Pasadena, CA 91125 USA
[4] TQC, University of Antwerp, Universiteislpein 1, 2610 Antwerpen, Belgium
* Corresponding author: yoan.leger@insa-rennes.fr

**Nonlinear interactions in coherent gases are not only at the origin of of bright and dark solitons and superfluids. At the same time, they give rise to phenomena such as multistability, which hold great promise for the development of advanced photonic and spintronic devices. In particular, spinor multistability in strongly-coupled semiconductor microcavities shows that the spin of hundreds of exciton-polaritons can be coherently controlled, opening the route to spin-optronic devices such as ultrafast spin memories, gates or even neuronal communication schemes. Here, we demonstrate that switching between the stable spin states of a driven polariton gas can be controlled by ultrafast optical pulses. While such a long-lived spin memory necessarily relies on strong and anisotropic spinor interactions within the coherent polariton gas, we also highlight the crucial role of nonlinear losses and formation of a non-radiative particle reservoir for ultrafast spin-switching.**

In coherent gases, the very nature of microscopic interactions often governs the macroscopic behavior of the ensemble. Bright solitons in optical fibers[1] result from attractive photon-photon interactions (or Kerr lensing effect) while $^4$He superfluidity[2] is caused by repulsive ones. A key goal of the scientific community is to precisely tailor these interactions either to understand more deeply the physics of quantum gases or to find novel applications for such quantum states of matter. In cold atom physics, a successful pathway consists in working with interacting mixtures of different particles, due to a different species or spin.[3][4][5] Pioneering experiments demonstrated the use of so-called Feshbach resonances to tune the interactions within the condensate from repulsive to attractive.[3][4]

A similar situation recently showed up in semiconductor physics. In high-Q-factor semiconductor microcavities, exciton-polaritons arise from the strong coupling between quantum well excitons and cavity photons.[6] While, the polariton photonic component provides direct optical access to the quantum fluid and allows condensation at cryogenic temperatures,[7] their excitonic component offers a varied set of solid-state microscopic interactions. The polariton polarization can be described with a pseudospin (up, down) corresponding both to the spin of radiative excitons ($\pm 1$) and to the light polarization ($\sigma\pm$). The Coulomb exchange origin of polariton-polariton interactions results in an important spinor anisotropy:[8] while co-polarized polaritons display repulsive interactions (the co-polarized interaction constant $\alpha_1$ is positive), the counter-polarized interaction is much weaker and presumably attractive due to van der Waals interactions that allow for stable multi-exciton complexes.[9][10] ($\alpha_2 < 0$ and $|\alpha_2| \ll |\alpha_1|$). These interactions have been successfully used for optical manipulation of polariton condensates[11][12][13][14] and are mandatory for the observation of superfluidity[15] or spinor excitations.[16]

As a result, polaritons are ideal systems for the investigation of multistability. After more than three decades of investigation, optical bistability in cold atom gases is now observed at the single photon scale.[17] However, higher order multistability in these systems remains extremely infrequent.[18][19] On the contrary, the field of polariton multistability has developed very recently with promising findings: bistability in planar polariton systems was first observed in 2004.[20] Multistability, using the spin degree of freedom, was proposed by Gippius[21] and demonstrated by Paraiso et al. in polariton traps.[22] Polariton spinor multistability has even been proposed as a mechanism to realize polariton-based neuronal communication. [23]

Spin multistability is an appealing pathway in spintronics as it enables the optical control of the spin state of a coherent gas composed of hundreds of particles interacting with a solid state environment. In view of realizing an optically-controlled spin memory, the spintronics community has mainly focused on magnetic materials and diluted magnetic semiconductors [24] down to the individual spin.[25] While the memory of carrier spins can hardly last longer than the nano- or microsecond range,[26] magnetic atom spins offer the advantage of ultrafast manipulation (sub-picosecond range) and of magnetic metastability. We will show here that polaritons, thanks to their strong spinor nonlinearities can be used as a competitive alternative for the realization of an ultrafast, compact, all-optical spin random access memory.

In this work, we demonstrate an all-optical ultrafast spin memory based on semiconductor polaritons. This phenomenon is the extension to the spinor case of a photonic device developed for telecom applications and known as "all-optical flip-flop memory" (AOFF). [27] Our spin memory can operate in both a bistable regime where the spin state of the driven polariton gas can switch from spin up to spin down and in a tristable one where a linear combination of up and down controlled by the polarization of the driving field is also accessible. The efficiency of this polariton switch confirms that incoherent processes such as nonlinear losses are at the origin of the phenomenon. The ultrafast switching times, limited by the resolution of our experimental setup, underline the strength of this process. Finally, we also evidence, in the tristable regime, the importance of the formation of a non-radiative reservoir that can affect the dynamics of the system.

# Results

### Power and Polarization dependent spinor multistability

We use trapped polaritons, confined in 3μm diameter engineered potentials of a semiconductor planar microcavity. As presented in Figure1 a, zero-dimensional polaritons display an atom-like spectrum with well separated discrete energy levels.[28][29][30] Confinement also reduces the influence of photonic disorder on the linewidth inhomogeneity.[31] The narrow linewidth (100 μeV) of the confined polariton ground state (GS) compared to the cw laser detuning of the excitation $\Delta_{cw} = E_{cw} - E_{GS}$ of 0.4 meV makes it possible to observe an exceptionally sharp multistability behavior thanks to the bistability condition: $\Delta_{cw} > \sqrt{3}\gamma$. The population of the driven polariton gas depends on both the laser-GS spectral overlap and the laser intensity. For a given polariton population ($n_\uparrow$, $n_\downarrow$), the Coulomb interactions are responsible for an energy shift of the two spin polariton states: $\Delta E_{\uparrow(\downarrow)} = \alpha_1 n_{\uparrow(\downarrow)} + \alpha_2 n_{\downarrow(\uparrow)}$. A feedback mechanism appears in the polariton population evolution and competes with the losses due to the finite polariton lifetime of about 15 ps in our sample. This interplay is sufficient to create multistability.[21] At a given excitation power (upper threshold) the feedback mechanism becomes dominant and one or both polariton states jump into resonance with the pump laser. The coherent population jumps to a high intensity regime (so called upper branch). When decreasing the excitation power the emission remains on the upper branch even below the upper threshold power, until one or both of the spin populations falls to the lower intensity branch (lower threshold). In this simple case, the decoupling between the hystereses of the spin populations $n_\uparrow$ and $n_\downarrow$ only depends on the values of the interaction constants and on the polarization of the pump laser, which can favor one or the other of the spin populations.

We previously demonstrated that other crucial mechanisms influence polariton multistability and can even dominate the feedback mechanism. First, biexciton formation and scattering towards the non-radiative reservoir, most likely composed of high momentum dark excitons, result in spin dependent non-linear losses within the polariton population. The damping of one spin population increases with the density of the opposite spin population. We observed a resonance of this mechanism 1.1 meV below the exciton resonance (Fig.1 c and d) with a linewidth of 0.3 meV. It is thus particularly favored in the present case when the polariton states come into resonance with the laser. This feedback mechanism was proved to play the central role in polariton multistability.[22][32] In addition, the formation of a non-radiative reservoir while pumping resonantly the system yields an unexpected additional blueshift of the polariton gas. This additional mechanism results in simultaneous upper thresholds for up and down polaritons and an extra increase of the dominant spin population when the minority population jumps down.

Second, the energy states of our polariton traps feature a linear polarization splitting of 20 µeV along the crystallographic axes of the sample, as reported in Ref. [22]. Despite its small value, this splitting has strong consequences on the multistability behavior. First, spin multistability becomes possible even under linearly-polarized excitation, while no spin population is seemingly favored.[32] If the linear polarization of the pump laser is not along the splitting axes, the polarization splitting results in a small rotation of the polariton spin in the Poincaré sphere (see Fig.1b and d). Depending on the orientation of the excitation polarization in the equatorial plane, up or down polaritons can be favored, modifying the multistability behavior. Obtaining multistability (and *a fortiori* spin switching) when placing the excitation polarization along one of the eigenaxes of the polarization is however quite difficult as it strongly favors or disadvantages the stability of the linear multistable branch.

Figures 2a and 2b show typical double hysteresis curves obtained when driving polaritons with a cw laser polarized linearly off the linear splitting axes. While upper thresholds coincide, the lower thresholds are decoupled. When one of the two spin populations falls down to the lower branch, the nonlinear losses greatly decrease for the other spin population, the intensity of which increases. This gives rise to a second hysteresis loop. The experimental conditions we used to obtain multistability are described in more details in methods. The yellow plot of Figure 2c) shows the typical result of a polarization sweep of the cw laser around the linear polarization of Figure 2 a and b, for a fixed excitation power. The power has been chosen to lie between the lower thresholds of the two polariton populations (see yellow bar in Fig. 2a). Before the sweep, the system is in the up state (polarization degree value +1, see Fig. 1b)). The circular polarization degree of the cw laser is then swept from +0.5 to -0.5 and backwards. Under these conditions it is possible to completely flip the spin orientation of the system by changing the pump polarization (Fig. 2c, yellow plot). We obtain a bistable system with two possible spin orientations up and down : the system behaves as a Schmitt trigger for spins. At this cw power only one population is supported on the upper branch.

A multistable system is obtained by choosing a higher excitation power (see blue bar in Fig. 2a and b). We stop the power scan between the two lower thresholds when the system spin is up (yellow line as before) and increase the excitation power again to enter the second hysteresis between the upper thresholds and the lower σ− threshold (blue bar in Fig. 2a). When placing the excitation power in the new hysteresis region (blue bar in Fig. 2a) and performing a polarization sweep, we obtain a tristable system [22](Fig. 2c, blue plot). The three stable states are +1 (only the up polariton population is on the upper branch), -1 (only the down polariton population is on the upper branch) and 0 (both populations are on the upper branch).

**Ultrafast bistable spin switching**

The AOFF operation consists in the use of trains of ultra-fast optical pulses of tailored polarization to switch our driven polariton gas to one or the other branch of the hysteresis curves of Figure 1c. Each pulse of the train can be either (σ+), (σ−) or linearly polarized, depending on the desired operation. A delay line controls the time difference between the two pulses. Due to their large bandwidth the laser pulses are always overlapping with the investigated polariton state. On the detection side the polariton emission is spatially separated into ($I_{\sigma+}$) and ($I_{\sigma-}$) and time resolved with a streak camera. From the two signals the circular polarization degree $\rho_s = (I_{\sigma+} - I_{\sigma-})/(I_{\sigma+} + I_{\sigma-})$ is obtained and used as readout value of the system.

Let us first focus on the bistable case, where we establish switching between σ− and σ+ with very high contrast and very fast switching times (see Fig.3). The cw power has to be set in the conditions of the yellow plot of figure 2c. The system can be initialized in any configuration (both spin components on the lower branch or one of them in the upper branch). For the sake of clarity, we assume here that at the beginning both polariton populations are on the lower bistability branch. The AOFF can be written by sending in a σ+ pulse for example. In this case the pulse instantly creates a large up polariton population through resonant excitation. This leads to a blue shift of this population that comes in resonance with the cw laser, placing the system in the multistable state +1.

The signal intensities and hence their ratio strongly depend on the GS energy and on the cw laser detuning. We measured ratios between 10:1 and 20:1. When expressed in polarization degree the contrast between the two memory states is extremely high. We observe a jump from +0.9 to -0.9, which means that the polariton population is either 95% spin up or 95% spin down polarized. The required pulse energies also strongly depend on these two parameters and were usually lying between 0.6 and 1.5 pJ for reliable switching operation. Considering the spot size of the pulsed laser, the switching energy can be lower than 15 fJ/μm.

The spin memory information remains stored as long as the cw laser drives the polariton gas. When a σ− pulse is sent in, a large down polariton population is created fast in the sample within the pulse duration. This has two main effects. The first one is the blue shift of the down polariton population; the down polariton population jumps to the upper branch. In addition, nonlinear losses become significant, favoring the decay of both polariton spin populations and the formation of the reservoir. The up state is red shifted and hence decoupled of the cw pump. The up population falls to the lower branch while the down population, due to the additional population provided by the pulse, remains on the upper bistability branch. The memory value is now -1 and remains so till a σ+ pulse is sent in again. Figures 3c and 3d demonstrate that the AOFF operation is reversible and that both polariton populations remain stable on their bistability branches for at least 12 ns (the repetition rate of our pulsed laser) with no preferential polarization. In fact this spin information remains stored without any time limitation as long as the polariton gas is driven by the cw laser.

**Tristable spin switching**

We now focus on the tristable spin switch, presented in Figure 4. In this case, the power of the linearly polarized cw laser is set at 0.9 mW, which is in the region where the system supports three spin states (see vertical blue lines in Fig. 2). The switching between +1 and -1 is driven by applying σ+ and σ− pulses in the same way as described previously. However, we need to use linearly polarized laser pulses to switch to 0. Note that, in such conditions, the pulse generates an excess in both populations which leads to strong nonlinear losses. However the induced blueshift is enough to raise both spin populations to the upper bistability branch. The system remains in this state until a σ+ or σ− pulse is sent in. In this regime, we demonstrate switching operations between circular (σ+ or σ−) and linear (Fig. 4a and b) and between circular and counter-circular (Fig. 4c and d) polarizations. We observe that these dynamics are characterized by a slow transient between the arrival of the pulses and the stabilization of the polariton population, resulting in some noise in the polarization degree during the first 300 ps of the switch operation (Fig. 4b and 4d).

**Numerical simulations**

The multistable behavior, including its dynamics can be qualitatively reproduced by the use of a set of nonlinear equation describing the evolution of a single polarization doublet state for excitons $\chi_{+(-)}$, photons $\varphi_{+(-)}$ and a non-radiative reservoir with a population n. This set of equations has been introduced in Ref.[32] of the main article and takes into account both the real and complex part of polariton-polariton interactions, i.e., both blueshifts and nonlinear losses. A reservoir induced blueshift is also considered (see Methods).

The numerical simulations are in good agreement with the observed power dependence of the polariton multistability (Fig. 5, left panel). The upper thresholds coincide for both spin populations, while the lower thresholds are decoupled and accompanied by a blueshift of the dominant spin population, opening a second hysteresis in the evolution. As for the polarization dependence (Fig. 5, right panel), we faithfully reproduce the merged hysteresis revealing three stable states when the excitation polarization is almost linear. The influence of the laser linear polarization direction during the polarization sweep is consistent with the experimental one reported in Ref. [32] and with the theoretical results of Ref. [33].

We also obtain a qualitative agreement on the multistability dynamics, as shown in Figure 6. In this case we account for the incoherence between the cw laser and the pulsed excitation by introducing a random phase term between the two sources. The presented calculations are averaged results over 50 indivudal simulation runs. We can reproduce the bistable (a) and tristable switches (b and c) in conditions close to the experimental ones. In the tristable case where the cw excitation power is larger, we observe an elongation of the reservoir decay time as a consequence of the efficient reservoir feeding due to biexciton formation. This affects the rise time of the polariton spin population as in the experiment. Despite the averaging over multiple runs, we could not prevent population relaxation oscillations occuring after the pulses. In addition, the reproducibility of the switch , equivalent to a bit error rate, is much better in the experiment than in the simulations (Fig. 6 c). This can be seen by the non-negligible remaining switched-off population and the strong ellipticity of the polariton state instead of a purely linear one. Other parameters would have to be introduced in the calculation in order to account for these effects such as a more complete modelization of the polariton states (photonic trap, higher-energy states), saturation effects and a microscopic decription of the biexciton resonance.

## Discussion

The demonstration of effective ultrafast switching of our polariton spin memory reveals that, despite the coherence of the system, polariton-polariton interactions can be used to control the spin state of a polariton population at will, with external optical pulses. This manipulation is only possible if incoherent processes such as nonlinear losses play the major role in the polariton dynamics as confirmed by our numerical simulations.

In the bistable operation regime, the switching time reveals the dynamics of the spinor polariton-polariton interactions: the rise time of the increasing spin population signal follows its energy-blueshift evolution time, and the decay time of the decreasing spin population signal trails the reservoir population mediated by biexciton formation. The inset offigure 3b , obtained with a higher temporal resolution, demonstrates a spin-flip time of about 5 ps, which is the temporal resolution of our streak camera. The ultimate limitation of the spin flip time is the time scale of the nonlinear spinor polariton-polariton interactions, which is expected to be much faster.[34][35][36]

In the tristable regime, whereas the decay of a given spin population is as fast as in the bistable case, the rise time of the majority spin population is around 300 ps. Such a long time is absolutely unexpected in purely polaritonic systems, which usually feature dynamics in the range of a few ps. In addition, the tristable window of the pump settings is very narrow and fluctuations tend to stabilize the system into the linear polarization state, which is the most stable. As a consequence, in order to switch to the σ+ and σ− states, pulse energies have to be higher than those used in the bistable case (between 4.1 pJ and 5.1 pJ). The higher powers of both cw pump and pulses favor the interaction between counter-polarized polaritons and the formation of a strong non-radiative reservoir, impacting the dynamics of the switch: When a pulse arrives, both spin populations are large enough so that the non-radiative reservoir forms efficiently due to the biexciton mediated process. This strong reservoir still contributes to the blueshift of both polariton spin populations. It can be sufficient to shift the polariton states out of resonance from the cw laser, producing a sharp drop in the polariton intensity just after the arrival of either laser pulse (linearly or circularly polarized). As the reservoir population decays, the resonance is progressively restored and the polariton population slowly increases until a steady state for the reservoir density is reached. The switch transient is thus an indirect measurement of the reservoir lifetime as confirmed by the numerical simulations.

In conclusion, we demonstrated here an ultrafast all-optical spin memory based on a semiconductor microcavity in the strong-coupling regime. The multistability dynamics allows us to reveal new features of spinor interactions within a polariton gas. In the bistable regime, the ultrafast switching time reveals the extremely efficient nonlinear losses between counter-polarized polariton spins. The slow dynamics of the tristable operation is a clear sign of the formation of a large non-radiative reservoir the interactions of which greatly affect the polariton gas spin state.

## Methods

**Multstability measurements**
The sample we used consists of a single $In_{0.04}Ga_{0.96}As$ quantum well placed at an antinode position in the middle of a GaAs/AlAs semiconductor microcavity. The Rabi splitting is 3.5 meV. The polariton traps are shallow mesas which consist of small local elevations (6 nm high and 3 µm in diameter) of the spacer layer of the microcavity. This leads to a local decrease of the microcavity resonance. The sample is placed in a liquid He continuous flow cryostat in transmission configuration, operating at 5K. The polariton gas is driven with a monomode cw Ti:sapphire laser focused with a f=1/1.2 camera objective providing a spot size of 18 µm (FWHM) at normal incidence. The here-presented results were obtained at a detuning between cavity and exciton mode of approximately +0.3 meV. We excited the sample with a linearly polarized cw laser which direction can be tuned with a half-wave plate. Polarization sweeps is achieved by rotating a quarter wave plate in the path of the cw laser. The laser detuning is $\Delta_{cw}$ = 0.4 meV above the ground state eigenenergy of the trapped polaritons. This energy is well below the eigenenergy of the first excited state (E1) which is separated from the ground state by approximately 0.8 meV. The zero in-plane momentum of the excitation guarantees maximum overlap with the ground state in momentum space and minimal overlap with the first excited state.[37] In this way we can neglect the contribution of E1 to the measured emission.

**Spin switching measurments**
The AOFF operation is achieved by adding to the experimental setup an extra excitation scheme composed of trains of 150 fs optical pulses of controlled polarization provided by a pulsed Ti-Sapphire laser coupled to a polarizing Michelson interferometer. The spot size of the pulses is 12 µm (FWHM)). To track the time evolution of the polariton spin populations, the σ+ and σ− polariton emissions are separated with a Wollaston prism and imaged simultaneously along the silt of a Streak camera with 5ps resolution.

**Equation set and numerical parameters**
We use the following set of equations:

$$\begin{cases} \dot{\chi}_{\uparrow\downarrow} = \left(-\left(\frac{\gamma_X}{2}+\beta|\chi_{\downarrow\uparrow}|^2\right)-i(\omega_X+\alpha_1|\chi_{\uparrow\downarrow}|^2+\alpha_2|\chi_{\downarrow\uparrow}|^2+\alpha_R n)\right)\chi_{\uparrow\downarrow}-i\frac{\Omega}{2}\varphi_{\uparrow\downarrow} \\ \dot{\varphi}_{\uparrow\downarrow} = \left(-\frac{\gamma_C}{2}-i\omega_C\right)\varphi_{\uparrow\downarrow}-i\frac{\delta e^{\pm 2i\theta}}{2}\varphi_{\downarrow\uparrow}-i\frac{\Omega}{2}\chi_{\uparrow\downarrow}+F_{\uparrow\downarrow} \\ \dot{n} = -\gamma_R n+\beta|\chi_{\downarrow\uparrow}|^2|\chi_{\uparrow\downarrow}|^2 \end{cases}$$

In these equations, $\alpha_1$, $\alpha_2$ and $\alpha_R$ are the copolarized, counter-polarized and reservoir induced interaction constants respectively. We use $\alpha_1$=10µeV. µm$^{-2}$, $\alpha_2$=-0.15 $\alpha_1$ and $\alpha_R$=0.04 $\alpha_1$ to reproduce as best as possible the observed multistability behavior. The ratio between co- and counter-polarized interactions is in agreement with the literature and we use a very small value for $\alpha_R$ to account for the small overlap between trapped polaritons and the non-confined non-radiative reservoir. The nonlinear loss constant $\beta$ is set to 0.015ps$^{-1}$ µm$^{-2}$. $\gamma_X$=0.05 ps$^{-1}$, $\gamma_C$=0.05 ps$^{-1}$ and $\gamma_R$=0.01 ps$^{-1}$ are the exciton, photon and reservoir damping constants respectively, in consistency with the set of parameters considered in Ref [32] except for the reservoir lifetime and interactions, which have been slightly modified. The exciton and photon energies ($\omega_X$ and $\omega_C$) are set to 0. The Rabi splitting $\Omega$ is 3.5 meV. The linear polarization splitting, attributed to photonic mechanisms is described by the energy constant $\delta$=20 µeV and the direction $\theta$ of the polarization axes. In the simulations we consider $\theta$=0 so that the polarization axes of the polariton doublet are H and V. $F_{\uparrow\downarrow}$ accounts for the laser excitation, including the cw laser as well as the pulse trains.

## Acknowledgements

The authors would like to thank Nikolay Gippius for fruitful discussions and Roger Rochat for permanent technical support. This work was supported by the Quantum Photonics NCCR (128792) and the Swiss National Science Foundation(135003) .

## Author contributions

R.C., T.K.P. and Y.L. conceived the experiment. R.C. and Y.L. constructed the experimental set-up. R.C. ran all the experiments. Y.L. and M. W. provided theoretical support. F.M.G. made the sample. M.P.O. and B.D. supervised the project. All coauthors contributed to the writing of the paper.

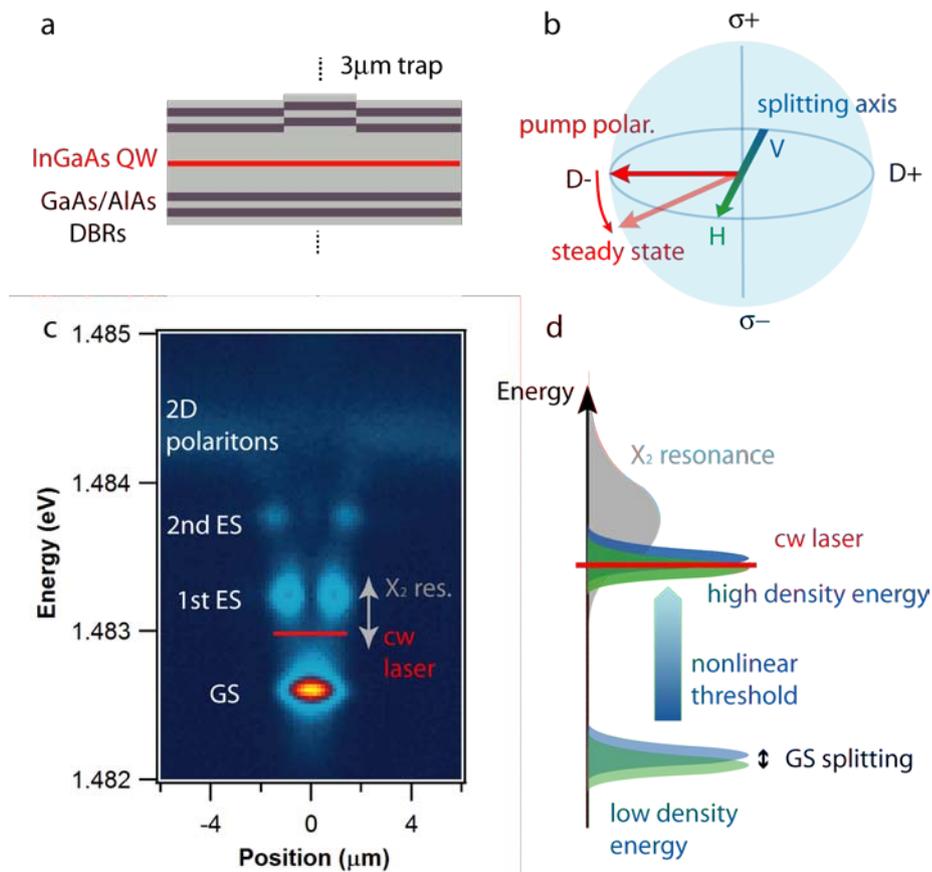

**Figure1 Quasi-resonant excitation of trapped polaritons.** (a) We use polaritons confined in 0D traps engineered in a GaAs/AlAs planar microcavity embedding an InGaAs quantum well. (b)To obtain multisatibilty conditions, the cw laser is blue detuned compared to the polariton ground state (GS) of the trap, close to the biexciton resonance ($X_2$ res.), which favors nonlinear losses. Excited states of the trap (ES) are nor populated. Increasing the intensity of the pump laser, a nonlinear threshold is reached and the polariton GS jumps into resonance with the laser (c and d). As the GS is a linear polarization doublet, the steady state of the polariton spin can be different from the injected spin If the cw laser polarization is not along one of the eigen axis of the polarization splitting, favoring one of the two spin populations (up or down).

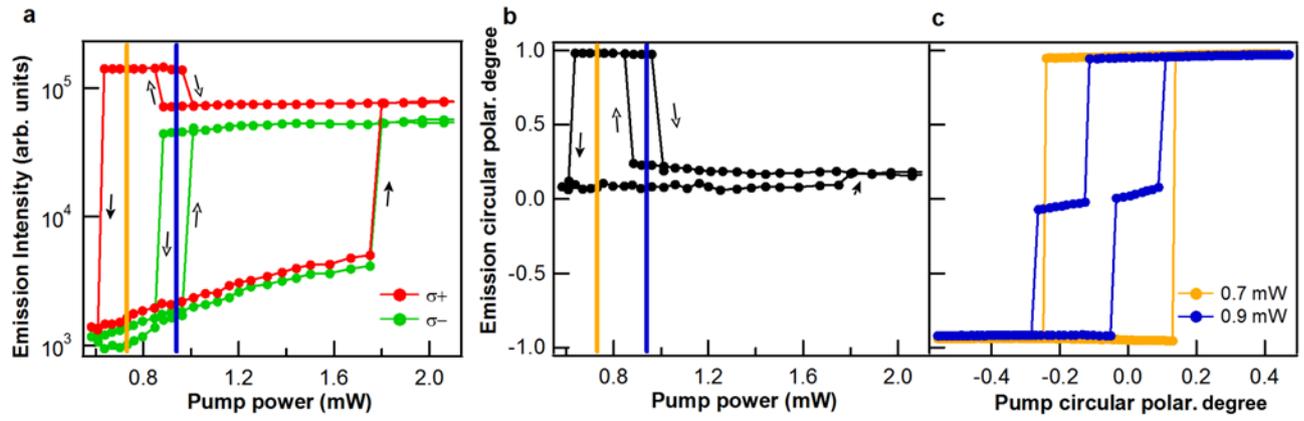

**Figure 2: Characterization of the polariton multistability.** For this purpose the ground state of the confined polaritons was excited with a cw laser at Δcw = 0.4 meV above the eigenenergy. For a) and b) the cw laser was linearly polarized and the excitation power was scanned from 0.6 mW to 2.2 mW and back to 0.6 mW. Panel a displays the bistability of the spin up (red) and spin down (green) polariton populations on a logarithmic scale (full arrows). When scanning upwards we observe the same upper threshold for both populations. On the way backwards however the two populations display independent thresholds. The multistability cycle is displayed by empty arrows. Panel b shows the polarization degree obtained from a). For c) yellow cycle (blue cycle): the system has been prepared in the polarization degree value +1 (yellow (blue) line in a)). Afterwards the polarization degree of the coherent emission has been measured while the polarization degree of the cw laser has been swept from +0.5 to -0.5 and backwards. We observe a hysteresis (tri-stable hysteresis) and the spinflip of the polariton population. The yellow (blue) vertical line indicates the cw laser parameters used for the memory measurements in the bistable (tristable) regime shown in Figure 3 (Figure 4).

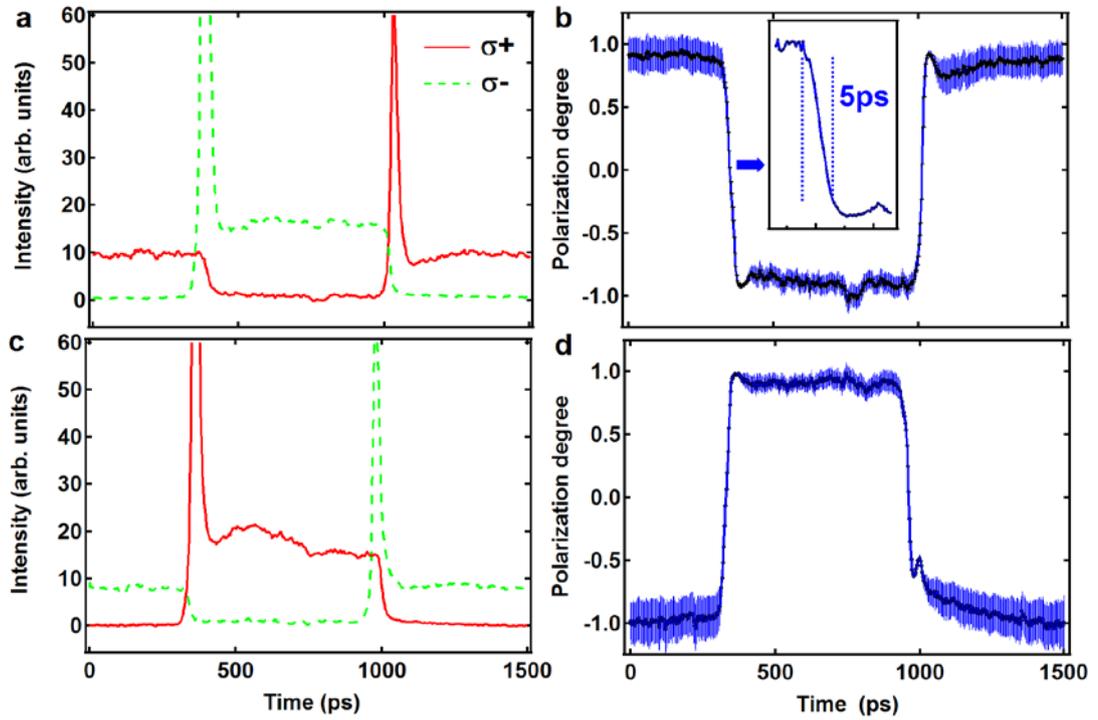

**Figure 3: Experimental polariton population and polarization rate dynamics in the bistable region.** The power of the cw laser was set at 0.7 mW (yellow plots of Figure 2). The emission coming from the two oppositely polarized polariton populations was separated and monitored with a streak camera. The time resolved σ+ and σ− signals in a) and c) clearly demonstrate switching behavior. The time resolved polarization degree (Iσ+ − Iσ−)/(Iσ+ + Iσ−) measurements in b) and d) demonstrate a very high switching contrast from +0.9 to -0.9 which is very close to the maximal values of +1 and -1. The inset in b) has been obtained with a higher temporal resolution demonstrating a switching time of about 5 ps for the memory. In c) and d) the order of the σ+ and σ− pulses has been reversed with respect to a) and b). In panels b and d, error bars on the polarization degree are calculated by using the the standard deviation of each circularly polarized intensity over the different phases of the switch.

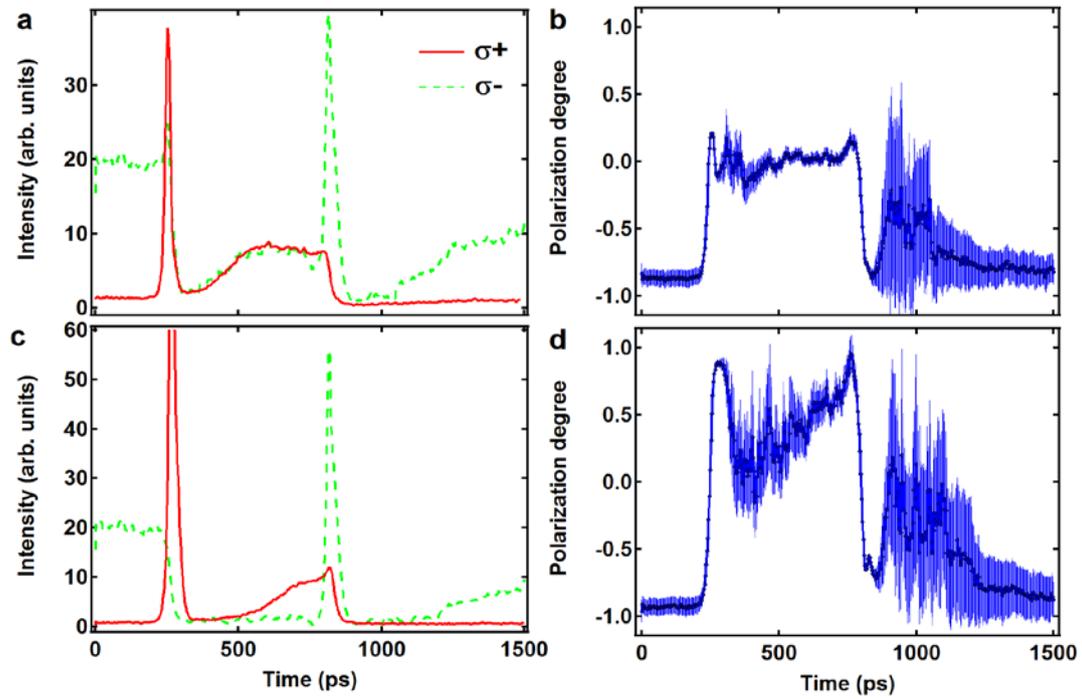

**Figure 4: Experimental polariton population and polarization rate dynamics in the tristable region.** The power of the cw laser was set at 0.9 mW (blue plots of Figure 2). The emission coming from the two oppositely polarized polariton populations was separated and monitored with a streak camera. a) Time-resolved switch from σ− to linear and back to σ−. b) The time resolved polarization degree (Iσ+ − Iσ−)/(Iσ+ + Iσ−) of the switching in a). c) Time-resolved switch from circular to counter-circular and back to circular polarization under the same conditions as in a). d) The time resolved polarization degree of the switching in c). In panels b and d, error bars on the polarization degree are calculated by using the the standard deviation of each circularly polarized intensity over the different phases of the switch.

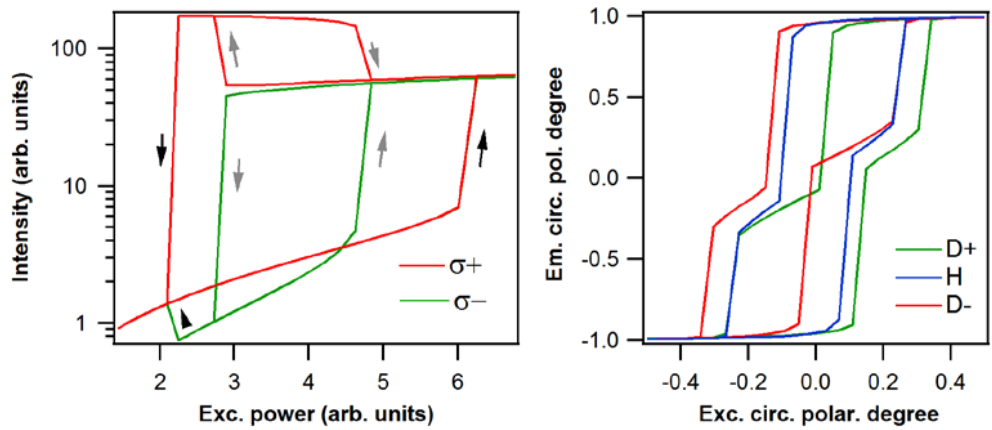

**Figure 5: Simulated power dependence and polarization dependence of the polariton multistability.** In the case of the power dependence (lef panel), the polarization of the pump laser is linear and is rotated by 20° with respect to the H axis so that the (σ+) population is favored. The polarization dependence (right panel) shows the change of the multistability curve with the direction of the linear polarization crossed during the polarization sweep of the pump laser. The excitation power is 4 for D+ and D- and 4.4 for H with respect to the power dependence.

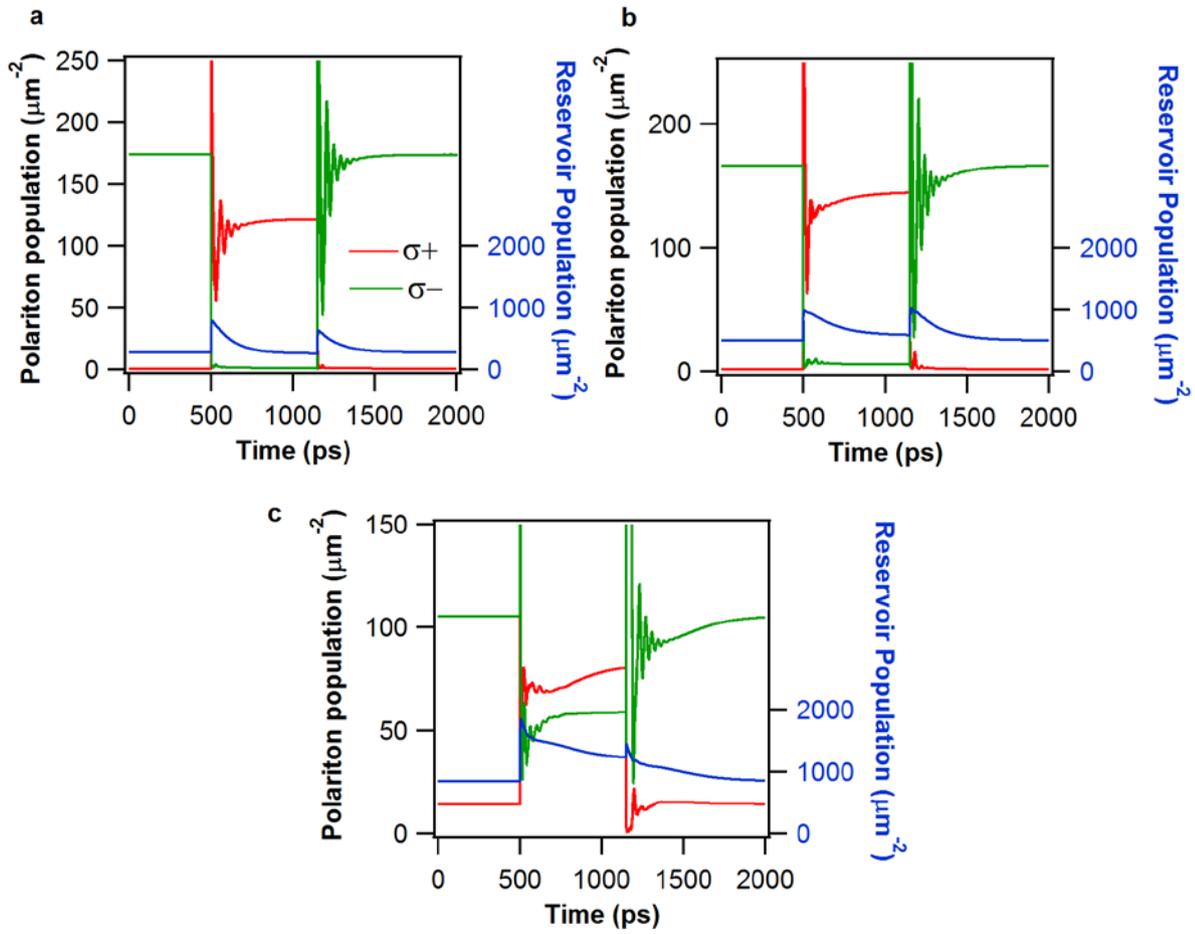

**Figure 6: Simulated switch experiments in the bistable and tristable case.** Panel a shows the bistable switch between σ+ and σ-. Panels b and c show the tristable switch: between σ+ and σ- in panel b and between linear polarization and σ- in c). The excitation power is 2.25 in the bistable case and 3.4 in the tristable case (see Figure 5).